\documentclass[fignum]{apa}
\usepackage[active]{srcltx} 
\usepackage{graphicx,bm,amsmath}
\usepackage{amsfonts}
\usepackage{graphicx}
\usepackage{natbib}
\usepackage{setspace}

\title{A factor mixture analysis model for multivariate binary data}
\author{Silvia Cagnone and Cinzia Viroli}
\affiliation{Department of Statistics, University of Bologna, Italy}

\abstract{The paper proposes a latent variable model for binary data coming
from an unobserved heterogeneous population. The heterogeneity is
taken into account by replacing the traditional assumption of
Gaussian distributed factors by a finite mixture of multivariate
Gaussians. The aim of the proposed model is twofold: it allows to
achieve dimension reduction when the data are dichotomous and,
simultaneously, it performs model based clustering in the latent
space. Model estimation is obtained by means of a maximum likelihood
method via a generalized version of the EM algorithm. In order to
evaluate the performance of the model a simulation study and two
real applications are illustrated.\\

}

\shorttitle{A factor mixture analysis model for multivariate binary data}
\doublespace
\begin{document}
\maketitle

\section{Introduction}
Observed binary variables are very common in behavioural and social
research.  Typical examples are those in which individuals can be
classified according to the fact that they agree/disagree to some
issues or to the fact that they can choose the right or wrong answer
in an educational test. Such binary variables are often supposed to
be indicators of one or more latent variables like, for instance,
ability or attitude. In the education field latent variables are
interpreted as `traits' so that usually factor models for binary
data are referred to as latent trait models (Lord and Novick, 1968; Moustaki, 1996).
These models can be viewed as a particular class of a more general family of latent variable models classified by Bartholomew and Knott (1999) according to the different nature of manifest and latent
variables. When both of them are continuous the latent variable model is the well-known classical factor analysis. When the former are continuous and the latter are discrete we have the latent profile analysis. If both are discrete we refer to the latent class analysis (Lazarsfeld,
1950, Goodman, 1974). With observed discrete variables and continuous latent variables, as in our situation, the latent variable model is called latent trait analysis.
The main feature of latent trait models is
that the conditional distribution of a complete
\textit{p}-dimensional set of responses of a given individual
(called response pattern) is specified as a function of the latent
variables. Each of the observed variables follows a Bernoulli
conditional distribution, whereas the latent variables are usually
assumed to be normally distributed. This assumption cannot be
appropriate when the observed data arise from some unobserved
sub-populations, so that the investigated population is not
homogeneous. In order to detect the potential classes or groups of
observations it can be more convenient to consider the latent space
as categorical, by performing thus the latent class analysis. According to this approach if a sample of
observations arises from some underlying sub-populations of unknown
proportions, its distributional form is specified in each of the
underlying populations and the purpose is to decompose the sample
into its latent classes. For this reason latent class is also
referred to as mixture-model clustering (McLachlan and Basford,
1988) or model-based clustering (Banfield and Raftery, 1993; Fraley
and Raftery, 2002). For an exhaustive description of latent class
analysis see, among the others, Vermunt and Magidson, (2003) and
Moustaki and Papageorgiou, (2005).

It is worth noting that the purposes of latent trait and latent class analysis
are different. Latent class analysis aims at performing clustering of
units whereas the aim of latent trait model is dimension reduction
(see Haberman, 1979, and Heinen, 1996, for a further discussion of similarities and differences between
the two approaches). When the interest focuses on both issues simultaneously a unified strategy has to be
considered.

With this purpose, Everitt (1988)
and Everitt and Merette (1990) proposed an extension of model based
clustering for mixed data in which the observed categorical
variables are generated according to the underlying latent variable
approach (Muth\'{e}n, 1984). Uebersax and Grove (1993) presented a latent trait model for the
analysis of agreement on dichotomous or ordered category ratings, in
which they assumed a finite mixture of $k=2$ distributions on the
univariate latent variable. They also imposed some restrictions on
the parameters of the mixture in order to guarantee model estimation
through a direct search optimization routine. More recently, a class of latent variable models, called factor mixture models, has been introduced (Yung; 1997, Muth\'{e}n and Shedden, 1999; Lubke and Muth\'{e}n, 2005) with the aim of measuring underlying
continuous constructs and simultaneously modeling population
heterogeneity by incorporating categorical and continuous latent variables. A drawback of this class of methods is that, for assumption, heterogeneity is exclusively ascribed to factor mean differences
across the latent classes, while factor variances and
covariances are held equal across classes. Montanari and Viroli (2010) proposed an
alternative way to deal with unknown heterogeneity by explicitly
assuming that the factors underlying a set of continuous observed
variables follow a mixture of multivariate Gaussians. In so doing,
heterogeneity is fully expressed by factor differences in mean and
variance components and the density of the observed variables proves
to be a mixture of $k$ heteroscedastic Gaussians, thus improving flexibility and
clustering performance.

 The aim of this work is to present an extension of heteroscedastic factor mixture analysis (Montanari and Viroli, 2010)
 for multivariate binary data. The proposed model contextually
performs dimension reduction and cluster analysis. Dimension
reduction is achieved by assuming that the data have been generated
by a factor model with continuous latent variables. In so
doing, we measure the `traits' of a latent trait model. Cluster
analysis is performed by assuming that the latent variables are
modelled as a multivariate Gaussian mixture, thus realizing a model
based clustering in the latent space. In this perspective the
proposed model can be also viewed as a more general latent trait
finite mixture model (Uebersax and Grove; 1993) by allowing
heteroscedastic and multivariate mixture components.

The paper is organized as follows. In the next Section the proposed
model is introduced and discussed. Section 3 is devoted to model
identifiability. A generalized EM algorithm for the estimation of
the model parameters is developed and illustrated in Section 4.
Section 5 presents the results of a simulation study. Two real
applications are finally illustrated.

\section{The model}
Let $\mathbf{y}=(y_{1},\ldots,y_{p})$ be a vector of $p$ observed
binary variables. We assume they are measurements of $q$ latent
variables, $\mathbf{z}=(z_{1},\ldots,z_{q})$ or, equivalently, the
associations among the observed variables are wholly explained by $q$
latent variables, with $q<p$. The relation between $\mathbf{y}$ and
$\mathbf{z}$ can be modelled through a monotone differentiable link
function, like the logit, as follows
\begin{equation}\label{eqn:logit}
\textrm{logit}(\pi_{j}(\mathbf{z}))=\lambda_{j0}+\sum_{r=1}^q
\lambda_{jr}z_{r}=\lambda_{j0}+{\boldsymbol \lambda}_{j}^T\textbf{z} \quad
j=1,\ldots,p
\end{equation}
where $\pi_{j}(\mathbf{z})=P(y_{j}=1|\mathbf{z})$.  The parameters
$\lambda_{j0}$ and ${\boldsymbol
\lambda}_{j}=(\lambda_{j1},\lambda_{j2},\ldots,\lambda_{jq})$ are
intercept and factor loadings, respectively, as in the classical
factor analysis. Alternatively to the logit, we could have referred
to the probit link function (Lord and Novick, 1968).\par The joint
density function of the response pattern
$\mathbf{y}=(1,0,0,\ldots,1)$ is given by
\begin{equation}\label{eqn:prima}
f(\mathbf{y}) =\int_{R^{q}} f(\mathbf{y}|\mathbf{z})f(\mathbf{z})d\mathbf{z}
\end{equation}
where $f(\mathbf{y}|\mathbf{z})$ is the conditional distribution of
$\mathbf{y}$ given $\mathbf{z}$ and $f(\mathbf{z})$ is the density
function of $\mathbf{z}$. As in the classical factor analysis, we assume that the association among the observed variables is wholly explained by the $q$ latent variables that is the conditional independence of $\mathbf{y}$ given $\textbf{z}$:
\begin{eqnarray}\label{eqn:sec}
f(\mathbf{y}|\mathbf{z})= \prod_{j=1}^p f(y_{j}|\mathbf{z})=
\prod_{j=1}^p
\pi_{j}(\mathbf{z})^{y_{j}}[1-\pi_{j}(\mathbf{z})]^{1-y_{j}},
\end{eqnarray}
that is, each $f(y_{j}|\mathbf{z})$ follows a Bernoulli
distribution. In the classical latent trait model $\mathbf{z}$ is
assumed to be multivariate normally distributed. Here, we
investigate the alternative assumption that the vector of latent
variables $\textbf{z}$ is distributed according to a finite mixture
of multivariate Gaussians
\begin{eqnarray}\label{eqn:mixture1}
f(\textbf{z})= \sum_{i=1}^{k}\tau_i \phi_{i}^{(q)} ({\boldsymbol \mu}_i,
{\boldsymbol \Sigma}_i),
\end{eqnarray}
where $\tau_i$ are the unknown mixing proportions, $\phi_{i}^{(q)}$
is the Gaussian density with vector mean $\boldsymbol \mu_i$ and covariance
matrix ${\boldsymbol \Sigma}_i$ of order $q$. This assumption implies that,
with respect to the latent variables, the overall population may be
heterogeneous and it is composed of $k$ distinct sub-groups, with
distribution defined by each Gaussian component. This allows to achieve one of the main aims of this work:
 the units or individuals may be classified on the basis of their factor values. In particular, clustering is obtained in a dimensionally reduced space defined by the latent traits.

More specifically, clustering is performed by a discrete
latent variable we implicitly introduce by modeling the factors
through a multivariate mixture of Gaussians. Since its role is to
\emph{allocate} each observation to one out of the $k$
sub-populations of the mixture, it is called \emph{allocation
variable}. The allocation variable is a $k$-dimensional random
variable denoted by $\textbf{s}=[s_i]_{i=1,\ldots,k}$ following a
multinomial distribution
\begin{eqnarray}\label{eqn:term2}
f(\textbf{s})= \prod_{i=1}^k\tau_i^{s_i},
\end{eqnarray}
and therefore $Pr(s_1=0,\ldots,s_i=1,\ldots,s_k=0)= \tau_i$.

Thus the proposed model involves two kinds of latent variables,
$\mathbf{z}$ and $\mathbf{s}$, which accomplish the different tasks
of dimension reduction and clustering, respectively.

The relation between observed and latent variables can be posed in a
hierarchical structure. Since $\textbf{s}$ is related to
$\mathbf{y}$ only through $\mathbf{z}$, the so-called
\emph{complete} density of the model
$f(\textbf{y},\textbf{z},\textbf{s})$, can be rephrased in the
following hierarchical form

\begin{eqnarray}\label{eq:hier}
f(\textbf{y},\textbf{z},\textbf{s})&=&f(\textbf{y}|\textbf{z})f(\textbf{z}|\textbf{s})f(\textbf{s}),
\end{eqnarray}
which allows to estimate the model by a generalized version of
the EM algorithm, that will be presented in the following.

\section{Model identifiability}

The identifiability of the model is crucial to obtain unique and consistent
parameter estimates. The model is not identified when it can be equivalently expressed by different parameterizations. The proposed model (\ref{eqn:logit}) can be reformulated in compact form as follows

\begin{equation}\label{eqn:logitC}
\textrm{logit}(\boldsymbol \pi(\mathbf{z}))=\boldsymbol \lambda_{0}+
{\boldsymbol\Lambda}\mathbf{z}
\end{equation}
where $\boldsymbol \pi(\mathbf{z})=\left( \pi_1(\mathbf{z}),\ldots, \pi_p(\mathbf{z})\right)$, $\boldsymbol \lambda_{0}$ is the $p$-dimensional vector of intercepts and
$\boldsymbol\Lambda$ is the $p\times q$ matrix of the factor loadings. Notice that (\ref{eqn:logitC}) is completely
indistinguishable from
$\textrm{logit}(\boldsymbol \pi(\mathbf{z}))=\boldsymbol \lambda_{0}+
\boldsymbol\Lambda\mathbf{M}\mathbf{M}^{-1}\mathbf{z}$, where $\mathbf{M}$
is an invertible square matrix of dimension $q \times q$. In order to avoid this ambiguity and to make
the model fully identified $q^2$ restrictions have to be imposed (because of the dimension of $\textbf{M}$).

One solution is the following. As in the classical factor analysis, we assume the factors are standardized,
\textit{i.e.} the mixture parameters must satisfy:
\begin{eqnarray}\label{eqn:standardized}
E(\textbf{z})&=&\sum_{i=1}^k\tau_i{\boldsymbol \mu}_i=\textbf{0},
\\\label{eqn:standardized2}
Var(\textbf{z})&=&\sum_{i=1}^k\tau_i\left({\boldsymbol \Sigma}_i+{\boldsymbol \mu}_i{{\boldsymbol \mu}}_i^\top\right)-
\left(\sum_{i=1}^k\tau_i{\boldsymbol \mu}_i\right)\left(\sum_{i=1}^k\tau_i{\boldsymbol \mu}_i\right)^\top
=\textbf{I}_q.
\end{eqnarray}

This implies that the factors are
uncorrelated but it is worth noting that they may be mutually
dependent, since for non-Gaussian random variables uncorrelatedness
does not imply independence and therefore potential nonlinear
relationships between the factors are preserved.
The previous assumptions introduce $q$ and $q(q - 1)/2$ constraints in the estimation of the
mean and covariance components, respectively. There are still $q^2-q(q+1)/2=q(q-1)/2$ restrictions to be imposed: in fact the model is still invariant under orthogonal rotations.
Thus, as proposed by J\"{o}reskog (1969), we introduce the further restrictions on the
loading matrix $\boldsymbol\Lambda$ by fixing $q(q-1)/2$ loadings equal to 0 (as rule of thumb the upper triangular of $\boldsymbol\Lambda$ is fixed to zero).

The previous constraints are necessary and sufficient to obtain the uniqueness of the model solution. A further identifiability aspect to be considered is the existence of the solution. For factor models it is guaranteed by the well-known Lederman's condition (1937), which relates the number of admissible
common factors $q$ to the number of observed variables $p$:
\begin{eqnarray}\label{eqn:ledcon}
k\leq\frac{1}{2}\{2p+1-\sqrt{8p+1}\}.
\end{eqnarray}

\section{Generalized EM algorithm}\label{EM}
In order to develop the EM algorithm for the proposed model a more compact notation must be defined.
Let $\textbf{\~{z}}$ denote the latent variables with an added column of ones, $\textbf{\~{z}}=(\textbf{1},\textbf{z})$, and $\tilde{\boldsymbol{\Lambda}}$
the matrix of dimension $p \times(q+1)$ which contains also the intercepts.
The model parameters are collectively
denoted by $\boldsymbol \theta=\left\{\boldsymbol{\tilde{\Lambda}},\boldsymbol \tau,\boldsymbol \mu,\boldsymbol \Sigma\right\}$ where $\boldsymbol \tau$ is the vector of $k$ weights, $\boldsymbol \mu=(\boldsymbol \mu_1,\ldots,\boldsymbol \mu_k)$ and $\boldsymbol \Sigma=(\boldsymbol \Sigma_1,\ldots,\boldsymbol \Sigma_k)$.
$\boldsymbol \theta$ can be efficiently estimated by the EM algorithm (Dempster
\textit{et al.}, 1977). Let
${\textbf{y}}=({\textbf{y}}_1,...,{\textbf{y}}_n)$ denote the
observed sample of size $n$. The EM algorithm consists of maximizing
the conditional expected value of the complete log-likelihood given
the observed data:
\begin{eqnarray}\label{eq:condexp}
\arg \max_{\boldsymbol \theta}E_{\textbf{z},\textbf{s}|\textbf{y},
\boldsymbol \theta'}\left[ \log \sum_{h=1}^n
f\left(\textbf{y}_h,\textbf{z}_h,\textbf{s}_h;\boldsymbol \theta\right)
\right].
\end{eqnarray}
Thanks to the decomposition in (\ref{eq:hier}), the problem
simplifies in evaluating the conditional expectation of the
logarithm of the three densities and the estimation algorithm has
the following structure:
\begin{enumerate}
\item Choose starting values $\boldsymbol \theta'=\left\{\tilde{\boldsymbol{\Lambda}}',\boldsymbol \tau',\boldsymbol \mu',\boldsymbol \Sigma'\right\}$.
\item Calculate:

\ \ \ \ \ \ \ \ \ \ - $\tilde{\boldsymbol{\Lambda}}$ which maximizes
$E_{\textbf{z}|\textbf{y}, \boldsymbol \theta'}\left[ \log
f(\textbf{y}|\textbf{z};\boldsymbol \theta) \right]$

\ \ \ \ \ \ \ \ \ \  - $\boldsymbol \mu$ and $\boldsymbol \Sigma$ which maximize
$E_{\textbf{z},\textbf{s}|\textbf{y}, \boldsymbol \theta'}\left[ \log
f(\textbf{z}|\textbf{s};\boldsymbol \theta) \right]$

\ \ \ \ \ \ \ \ \ \  - $\boldsymbol \tau$ which maximizes
$E_{\textbf{s}|\textbf{y}, \boldsymbol \theta'}\left[ \log
f(\textbf{s};\boldsymbol \theta) \right]$

\ \ \ \ \ \ \ \ \ \  - Set $\boldsymbol \theta'=\boldsymbol \theta$
\item If convergence is not achieved return to step 2. Convergence is attained when the

\ \ \ \ \ \ \ \ \ \ \ \  change in the observed data log-likelihood increases by less than a fixed $\epsilon$.
\end{enumerate}

\medskip

Since all the integrals involved in the expectation step cannot be
analytically solved, they are approximated by a weighted sum over a
finite number of points with weights given by the Gauss-Hermite
quadrature points (see, among the others, Straud and Sechrest, 1966
and Bock and Aitkin, 1981). Thanks to this approximation, estimates
for the mixture weights, means, and variance matrices can be
obtained in closed form. On the contrary, an analytical estimator
for the parameters contained in $\tilde{\boldsymbol \Lambda}$ cannot be derived but an iterative
Newton-Raphson procedure is needed. This leads to a generalized
version of the EM algorithm (GEM, see McLachlan and Krishnan, 2008)
whose E and M steps are described in the following.

\subsection{E-step}
In order to compute the conditional expected value of the complete
log-likelihood given the observed data, we need to determine the
conditional distribution of the latent variables given the observed
data on the basis of provisional estimates of $\boldsymbol \theta$,
$\boldsymbol \theta'$:
\begin{eqnarray}\label{eqn:EMstep}
f\left(\textbf{z},\textbf{s}|\textbf{y};\boldsymbol \theta'\right)=f\left(\textbf{z}|\textbf{y},\textbf{s};\boldsymbol \theta'\right)
f\left(\textbf{s}|\textbf{y};\boldsymbol \theta'\right).
\end{eqnarray}
Using Bayes' rule, the first term of the previous expression is
given by
\begin{eqnarray}\label{eqn:cond1}
f\left(\textbf{z}|\textbf{y},s_i=1;\boldsymbol \theta'\right)=\frac{f\left(\textbf{z}|s_i=1;\boldsymbol \theta'\right)f(\textbf{y}|\textbf{z};\boldsymbol \theta')}
{f\left(\textbf{y}|s_i=1;\boldsymbol \theta'\right)},
\end{eqnarray}
where $f\left(\textbf{z}|s^{(i)}=1\right)$ has the multivariate
Gaussian density with vector mean $\boldsymbol \mu_i$ and covariance matrix
${\boldsymbol \Sigma}_i$ and $f\left(\textbf{y}|\textbf{z}\right)$ is given
in expression (\ref{eqn:sec}). However
$f(\textbf{y}|s_i=1;\boldsymbol \theta')$ cannot be expressed in closed form
and must be approximated. Among the possible approximation methods,
Gauss-Hermite quadrature points are used here:
\begin{eqnarray}\label{eq:quad}
f(\textbf{y}|s_i=1;\boldsymbol \theta')&=&\int
f(\textbf{z}|s_i=1;\boldsymbol \theta')f(\textbf{y}|\textbf{z};\boldsymbol \theta')d\textbf{z}\\
\nonumber &\cong& \sum_{t_1=1}^{T_1}\ldots\sum_{t_q=1}^{T_q}w_{t_1}
\ldots
w_{t_q}f\left(\textbf{y}|\sqrt{2}{\boldsymbol \Sigma}_i^{1/2}\textbf{z}_t+{\boldsymbol \mu}_i;\boldsymbol \theta'\right)
\end{eqnarray}
where $w_{t_1},\ldots,w_{t_q}$ and
$\textbf{z}_t=(z_{t_1},\ldots,z_{t_q})^\top$ represent the weights
and the points of the quadrature respectively.

The second density of expression (\ref{eqn:EMstep}) is the posterior
distribution of the allocation variable $\textbf{s}$ given the
observed data which can be computed as posterior probability:
\begin{eqnarray}\label{posterior}
f(s_i=1|\textbf{y};\boldsymbol \theta')=\frac{f(s_i=1;\boldsymbol \theta')
f(\textbf{y}|s_i=1;\boldsymbol \theta')} {\sum_{i=1}^k f(s_i=1;\boldsymbol \theta')
f(\textbf{y}|s_i=1;\boldsymbol \theta')}.
\end{eqnarray}

\subsection{M-step}
The optimization step (a) of the algorithm consists in evaluating
and maximizing:
\begin{eqnarray}\label{eq:Mstep1}
E_{\textbf{z}|\textbf{y}; \boldsymbol \theta'}\left[ \log
f(\textbf{y}|\textbf{z};\boldsymbol \theta) \right]=\int\log
f(\textbf{y}|\textbf{z};\boldsymbol \theta)f(\textbf{z}|\textbf{y};\boldsymbol \theta')d\textbf{z}
\end{eqnarray}
with respect to ${\tilde{\boldsymbol\lambda}}_j=(1,\lambda_{j1},\ldots,\lambda_{jq}$) with $j=1,\ldots,p$,
where $f(\textbf{z}|\textbf{y};\boldsymbol \theta')=\sum_{i=1}^k
f(s_i=1|\textbf{y};\boldsymbol \theta')f\left(\textbf{z}|\textbf{y},s_i=1;\boldsymbol \theta'\right)$. Notice that $f(\textbf{y}|\textbf{\~{z}};{\boldsymbol \theta})=f(\textbf{y}|\textbf{z};{\boldsymbol \theta})$.
Let $S_0({\tilde{\boldsymbol\lambda}}_j,\textbf{y}|\textbf{\~z})$ be the derivative
with respect to $\tilde{\boldsymbol\lambda}_j$ of the log-density in
(\ref{eq:Mstep1}):
\begin{eqnarray*}
S_0({\tilde{\boldsymbol\lambda}}_j,\textbf{y}|\textbf{\~z})=\frac{\partial \log
f({\textbf{y}}|\textbf{\~z};{\boldsymbol \theta})}{\partial
{\tilde{\boldsymbol\lambda}}_j}=y_j\textbf{\~z}-\frac{\exp({\tilde{\boldsymbol\lambda}}_j^\top\textbf{\~z})\textbf{\~z}}
{1+\exp({\tilde{\boldsymbol\lambda}}_j^\top\textbf{\~z})}.
\end{eqnarray*}
Then the expected score function with respect to the parameter
vector $\tilde{\boldsymbol\lambda}_j$
\begin{eqnarray*}
\sum_{i=1}^k f(s_i=1|\textbf{y};{\boldsymbol \theta'})\int
S_0({\tilde{\boldsymbol\lambda}}_j,\textbf{y}|\textbf{\~z})
f\left(\textbf{\~z}|\textbf{y},s_i=1;\boldsymbol \theta'\right)d\textbf{\~z}=0
\end{eqnarray*}
can be evaluated by approximating the integrals with
Gaussian-Hermite quadrature points:
\begin{eqnarray}\label{eq:quad2}
&& \int S_0({\tilde{\boldsymbol\lambda}}_j,\textbf{y}|\textbf{\~z})
f\left(\textbf{\~z}|\textbf{y},s_i=1;\boldsymbol \theta'\right)d\textbf{\~z}\\
&& \cong \frac{1}{f(\textbf{y}|s_i=1;\boldsymbol \theta')}
\sum_{t_1=1}^{T_1}\ldots\sum_{t_r=1}^{T_q}w_{t_1}\ldots
w_{t_q}S_0({\tilde{\boldsymbol\lambda}}_j,\textbf{y}|\textbf{\~z}_t^*)f\left(\textbf{y}|\textbf{\~z}_t^*;\boldsymbol \theta'\right)
\end{eqnarray}
where $\textbf{\~z}_t^*=(1,\textbf{z}_t^*)$ and
$\textbf{z}_t^*=\sqrt{2}{\boldsymbol \Sigma}_i^{1/2}\textbf{z}_t+\boldsymbol \mu_i$. The
approximate gradient offers a non-explicit solution for the not null elements of the
parameter vector $\tilde{\boldsymbol \lambda}_j$, whose estimates can be obtained by
applying a constrained matrix in order to take into account the identifiability conditions on $\boldsymbol\Lambda$ (see, for major details, Tsonaka and Moustaki, 2007). The estimation problem can be solved by nonlinear optimization methods such as the Newton-type
algorithms (Dennis and Schnabel, 1983).

\noindent The optimization step (b) of the algorithm consists in
optimizing:
\begin{eqnarray}\label{eq:Mstep2}
E_{\textbf{z},\textbf{s}|\textbf{y}; \boldsymbol \theta'}\left[ \log
f(\textbf{z}|\textbf{s};\boldsymbol \theta) \right]=\sum_{i=1}^k\int \log
f(\textbf{z}|\textbf{s};\boldsymbol \theta)
f(\textbf{z},\textbf{s}|\textbf{y}; \boldsymbol \theta')d\textbf{z}
\end{eqnarray}
with respect to $\boldsymbol \mu_i$ and $\boldsymbol \Sigma_i$. By substituting
$f(\textbf{z},\textbf{s}|\textbf{y};
\boldsymbol \theta')=f(s_i=1|\textbf{y};\boldsymbol \theta')f(\textbf{z}|\textbf{y},s_i=1;\boldsymbol \theta')$
in the previous expression, the estimates of the new Gaussian
mixture parameters in terms of previous parameters $\boldsymbol \theta'$ are
\begin{eqnarray*}
{\boldsymbol \mu}_i&=&\frac{f\left(s_i=1|\textbf{y};\boldsymbol \theta'\right)E[\textbf{z}|s_i=1,\textbf{y};\boldsymbol \theta']}
{f \left( s_i=1|\textbf{y};\boldsymbol \theta' \right) },\\
{\boldsymbol \Sigma}_i&=&\frac{f\left(s_i=1|\textbf{y};{\boldsymbol \theta}'\right)\left(E[\textbf{zz}^\top|s_i=1,\textbf{y};{\boldsymbol \theta}']
-{\boldsymbol \mu}_i{\boldsymbol \mu}_i^\top\right)}
{f\left(s_i=1|\textbf{y};\boldsymbol \theta'\right)},
\end{eqnarray*}
where the first and second conditional moments,
$E[\textbf{z}|s_i=1,\textbf{y};\boldsymbol \theta']$ and
$E[\textbf{zz}^\top|s_i=1,\textbf{y};\boldsymbol \theta']$, can be computed
through the Gauss-Hermite quadrature points, similarly to
(\ref{eq:quad}) and (\ref{eq:quad2}). In order to take into account the identifiability conditions given in
(\ref{eqn:standardized}) and (\ref{eqn:standardized2}), the following scaling (\ref{eqn:scaling}) and centering (\ref{eqn:centering}) transformations are performed at each iteration:
\begin{eqnarray}\label{eqn:scaling}
&&\boldsymbol\Sigma_i \rightarrow (\textbf{A}^{-1})^\top\boldsymbol\Sigma_i\textbf{A}^{-1}, \ \ \ \boldsymbol\mu_i \rightarrow (\textbf{A}^{-1})^\top \boldsymbol \mu_i \\ \label{eqn:centering}
&&\boldsymbol\mu_i \rightarrow \boldsymbol\mu_i-\sum_{i=1}^k\tau_i\boldsymbol \mu_i,
\end{eqnarray}
where $\textbf{A}$ is the Cholesky decomposition matrix of $Var(\textbf{z})$.

Finally, the estimates for the weights of the mixture in step (c)
can be computed by evaluating the score function of
$E_{\textbf{z},\textbf{s}|\textbf{y}; \boldsymbol \theta'}[\log
f(\textbf{s};\boldsymbol \theta)]$ from which
\begin{eqnarray*}
{\tau}_i=f \left( s_i=1|\textbf{y};\boldsymbol \theta' \right).
\end{eqnarray*}

The algorithm has been implemented in R code and is available from
the authors upon request.

\section{Model selection}

The proposed model is characterized by two unknown dimensional quantities, the number of factors, $q$, and the number of groups, $k$. A possible way to perform model selection is to simultaneously choose $q$ and $k$ on the basis of some information criteria.
However, this procedure would require the estimation of all possible combinations of models with $q$ and $k$ varying in a range of admissible values. This would imply a high computational effort, especially when $q$ increases. A possible alternative procedure is inspired by the model assumption that the factors explain all the associations among the observed variables (\emph{i.e.} conditional independence assumption). This procedure represents a more computationally efficient solution in a forward selection strategy. In more details, it consists of two subsequent steps.

\subsection{Choice of the number of factors}

First, the most parsimonious models, with $q=1$ and $k$ varying from 1 to a maximum fixed value $K_{\textrm{max}}$, are fitted. If a single factor is not sufficient to explain the associations among the items (for at least one value of $k$), $q$ is increased by one until this condition is satisfied.
A possible measure to evaluate if the associations are completely taken into account by $q$ factors, is based on the so-called \emph{bivariate
residuals} (Bartholomew and Knott, 1999, Bartholomew \textit{et
al.}, 2002). They quantify the discrepancies between observed and
expected frequencies for each bivariate marginal frequency
distribution. Any large discrepancies will suggest that the model
does not fit the data well. As a rule of thumb a residual greater than 4 is considered large, as suggested by Bartholomew et al. (2002).

The bivariate residual based criterion does not represent the only possible measure considered in the goodness of fit literature for latent trait models. Alternatively classical statistical tests, like the likelihood ratio
($LR$) and the Pearson chi square ($GF$) can be used. However, unlike the
classical tests, the bivariate residuals are not affected by the
sparseness problem typical of categorical data, as it could occur with several binary items (see, for major
details, Reiser, 1996, Bartholomew and On Leung, 2002).

\subsection{Choice of the number of components}
Once $q$ has been selected, the number of components $k$ can be chosen according to the classical information criteria, such as the well-known BIC (Schwarz, 1978) and AIC (Akaike, 1974). This choice is very important because it is assumed that the number of components coincides with the number of groups. Therefore, the proposed model performs classification of units into $k$ groups in the latent space.

\section{Simulation study}
The effectiveness of the proposed model is first evaluated on a simulated study in order to analyze the goodness of fit and the classification performance.
Six different simulation designs have been implemented by considering $q=1,2,3$ and $k=2,3$. A number of
 $200$ samples with $n=300$ units have been generated for the different six experimental conditions.

\begin{figure}[t]
\centering
\includegraphics[height=100mm]{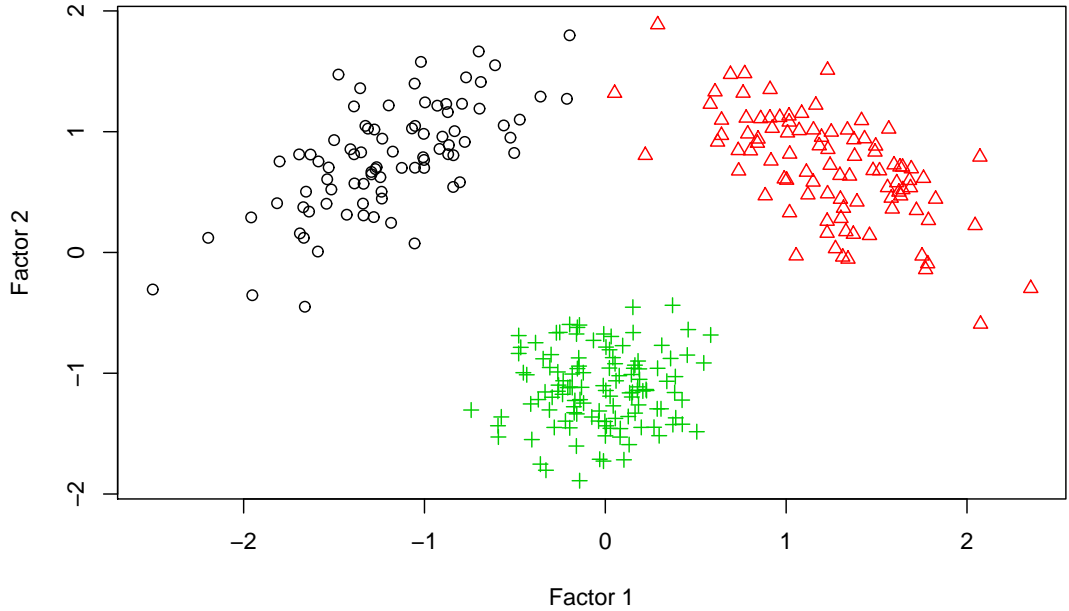}
\caption{Scatterplot of the factor scores distinguished by group.}
\label{factor}       
\end{figure}

In each experiment, a set of $p=10$ binary variables have been simulated according to the following factor loadings. The intercepts $\boldsymbol\lambda_{0}$ have been randomly generated within the range $[-3,3]$ and factor loadings $\boldsymbol\Lambda$ have been randomly generated within the interval $[2,4]$ in the case of $q=1$ and for $q>1$ within $[0,0.5]$ and $[2,4]$ for different subsets of items so that a quasi simple structure is produced. The factor parameters have been chosen in order to have quite well-separated and standardized factors as required for the identifiability of the model. For instance, with reference to the simulation design, $q=2$ and $k=3$, the two factors are modelled by a mixture of three multivariate Gaussians with mean and covariance components equal to
\begin{eqnarray*}
\mu_1=[-1.19 \ \ \  0.77], \ \ \ \ \ \ \ \  \mu_2=[1.20 \ \ \ 0.76], \ \ \ \ \ \ \ \ \mu_3=[-0.01 \ \ \ -1.15]
\end{eqnarray*}
\begin{eqnarray*}
\boldsymbol \Sigma_1=\left[
\begin{array}{rr}
0.17 & \ 0.08 \\
 0.08 & 0.14
\end{array}
\right], \ \
\boldsymbol \Sigma_2=\left[
\begin{array}{rr}
0.16 & \ -0.08 \\
 -0.08 & 0.12
\end{array}
\right], \ \ \boldsymbol \Sigma_3=\left[
\begin{array}{rr}
0.10 & \ -0.01 \\
 -0.01 & 0.09
\end{array}
\right]
\end{eqnarray*}
The weights of the mixtures have been fixed equal to $\tau_1=0.3$, $\tau_2=0.3$ and $\tau_3=0.4$.
Figure \ref{factor} shows the scatterplot of the two factor scores distinguished by group for one of the $200$ simulated samples.

For each sample within each experimental design, the previously described GEM-algorithm has been run by initializing the factor loadings with the conventional latent trait analysis solution (with one group). The remaining starting values for the model parameters have been randomly chosen. Eight quadrature points per each dimension have been chosen for the integral approximation.

Table \ref{tabella1} reports the model selection results obtained following the procedure previously described.
Columns refer to the different simulation designs.  The first three rows contain results about the choice of the
number of factors. For each experiment an increasing number of factors has been estimated until the percentage of samples (out of 200 generated samples) with low bivariate residuals is greater than $95\%$. Results show that the correct number of factors is selected in all the experimental conditions.

The remaining rows of the table relates to the choice of $k$, with $q$ given in the previous step. In particular, the percentages with which each fitted model is selected according to BIC and AIC are indicated.
According to the BIC the best number of groups is always underestimated with the exception of the first column ($q=1$ and $k=2$) where the true number of groups is chosen most of times. On the contrary, the AIC correctly suggests $k$ in
all the six situations. This is due to the fact that
the behaviour of the BIC is more conservative than that
of AIC because the former is characterized by a heavier penalty term which
depends on the sample size and could favour simpler
models. Therefore the less restrictive AIC seems to be more appropriate in our situation because of the complexity of the model.

\begin{table}[ht]
\caption{\label{tabella1} Model selection results for 6 different simulation designs. In the first three rows the percentages of samples (out of 200 generated samples) with low bivariate residuals are reported. In the last rows of the table the percentages with which each fitted model is selected according to BIC and AIC are indicated.}
\centering
\begin{tabular}{c|cc|cc|cc}
 & \multicolumn{6}{c}{\textrm{True model specification}} \\
  \hline
 &  \multicolumn{2}{c|}{$q=1$} & \multicolumn{2}{c|}{$q=2$} & \multicolumn{2}{c}{$q=3$}\\
 & \ \ $k=2$ \ \ & \ \ $k=3$ \ \ & \ \ $k=2$ \ \ & \ \ $k=3$ \ \ & \ \ $k=2$ \ \ & \ \ $k=3$ \ \ \\
 \hline
 & & & & & &\\
 $q=1$ & 98\% & 96\% & 0\% & 0\% & 0\% & 0\%\\
 $q=2$ & -- & -- & 100\% & 100\% & 0\% & 0\%\\
 $q=3$ & -- & -- &  -- & -- &  100\% & 100\% \\
 & & & & & &\\
 \multicolumn{1}{l|}{ \textrm{BIC} \ \ \ \ \ \ \ \ \ \ \ }& & & & & &\\
 $k=1$ & 38\% & 20\% & 82\% & 4\% & 96\% & 31\%\\
 $k=2$ & 43\% & 51\% & 18\% & 69\% & 4\% & 68\% \\
 $k=3$ & 16\% & 27\% & 0\% & 27\% & 0\% & 1\% \\
 $k=4$ & 3\% & 2\% & 0\% & 0\% & 0\% & 0\%\\
 & & & & & &\\
\multicolumn{1}{l|}{ \textrm{AIC} \ \ \ \ \ \ \ \ \ \ \ }& & & & & &\\
 $k=1$ & 27\% & 8\% & 22\% & 0\% & 30\% & 0\%\\
 $k=2$ & 42\% & 41\% & 50\% & 22\% & 54\% & 40\%\\
 $k=3$ & 24\% & 43\% & 22\% & 56\% & 12\% & 47\%\\
 $k=4$ & 6\% & 8\% & 6\% & 22\% & 4\% & 13\%\\
 \hline
\end{tabular}
\end{table}

Table \ref{tabella2} reports the means and the root mean square errors (rmse) for the previously described model specification $q=2$ and $k=3$. The factor loading estimates are quite accurate, even if a major precision could be obtained by increasing the sample size and the number of quadrature points. In order to
measure the classification performance of the proposed model, the
misclassification error between the true class membership and the
posterior classification of the estimated model obtained by equation
(\ref{posterior}) has been computed. The misclassification error
mean is 0.131 (with standard error of 0.009), thus indicating that
the 86.9\% of units are generally correctly classified.

\begin{table}[ht]
\caption{\label{tabella2} In the left part of the table, the true thresholds and
factor loadings of the simulated experiment are reported. The middle
part of the table contains the means of the thresholds and factor
loading estimates across the $200$ samples. In the last part of the table the corresponding root mean square errors are reported.}
\centering
\begin{tabular}{rrr|rrr|rrr}
  \hline
$\boldsymbol \lambda_0$ & $\boldsymbol \lambda_1$ & $\boldsymbol \lambda_2$ & \multicolumn{1}{c}{$\hat{\boldsymbol \lambda}_0$} & \multicolumn{1}{c}{$\hat{\boldsymbol \lambda}_1$} & \multicolumn{1}{c}{$\hat{\boldsymbol \lambda}_2$} & \multicolumn{1}{|r}{$\textrm{rmse}_0$} & $\textrm{rmse}_1$ & $\textrm{rmse}_2$\\
  \hline
0.45 & 2.16 & 0.00 & 0.49 & 2.83 & 0.00 & 0.42 & 1.02 & 0.00 \\
-0.99 & 3.21 & 0.26 & -1.04 & 4.35 & 0.29 & 0.71 & 1.77 & 0.59 \\
0.11 & 2.06 & 0.43 & 0.14 & 2.52 & 0.49 & 0.33 & 0.77 & 0.33 \\
-1.09 & 3.48 & 0.16 & -1.17 & 4.70 & 0.21 & 0.74 & 1.83 & 0.55 \\
-1.57 & 2.86 & 0.19 & -1.62 & 3.62 & 0.26 & 0.66 & 1.31 & 0.51 \\
-0.35 & 0.03 & 2.25 & -0.40 & -0.03 & 2.64 & 0.30 & 0.34 & 0.62 \\
0.23 & 0.07 & 2.22 & 0.23 & 0.05 & 2.62 & 0.27 & 0.35 & 0.70 \\
0.33 & 0.50 & 2.42 & 0.32 & 0.56 & 2.83 & 0.31 & 0.39 & 0.76 \\
0.35 & 0.47 & 2.25 & 0.36 & 0.53 & 2.73 & 0.28 & 0.39 & 0.85 \\
1.47 & 0.44 & 3.68 & 1.69 & 0.60 & 5.19 & 0.84 & 0.98 & 2.16 \\
   \hline
\end{tabular}
\end{table}

\section{Data Analysis}
\subsection{Example 1: Attitude towards Abortion}
This dataset has been extracted from the 1996 British Social
Attitudes Survey (Knott \textit{et al.}, 1990; McGrath and Waterton,
1986). Binary responses to four out of seven items concerning
attitude to abortion are given for $379$ individuals. The items
investigate circumstances under which an individual would consider
that an abortion should be allowed under law. The four circumstances
are:

\begin{itemize}
  \item $y_1$: the woman decides on her own that she does not.
  \item $y_2$: the couple agree that they do not wish to have a child.
  \item $y_3$: the woman is not married and does not wish to marry the man.
  \item $y_4$: the couple cannot afford any more children.
\end{itemize}

Possible responses are coded as 1 for `agree' and 0 for `not agree'.
Previous analysis (Bartholomew \textit{et al.}, 2002) on this data
suggested the presence of two classes in which the response patterns
can be grouped. The two classes can be easily interpreted as
`conservative' and `not conservative' attitude to abortion,
respectively. On the same data, latent trait analysis performed well
and highlighted that all the four items are good indicators of a
general factor summarizing the attitude towards abortion.

With the proposed model we aim at simultaneously performing a latent
trait analysis and grouping the response patterns of the dataset
into meaningful classes. To this purpose, a first numerical study
has been conducted in order to select among different models with
$k$ varying from 1 to 3. As far as the number of factors $q$ is
concerned, due to the Lederman's condition (\ref{eqn:ledcon}), given $p=4$ items only one factor can be estimated in an exploratory context. Moreover, $500$ different starting values for
the EM-algorithm have been considered and the best model in terms of
number of groups has been chosen according to the AIC criterion
(although in this case BIC leads to the same choice). Coherently
with the previous results on this data, the information criteria
suggest $k=2$.

\begin{table}[ht]\caption{\label{ab1} Threshold and loading estimates, with bootstrap standard errors in brackets, attitude to abortion.}
\centering
\begin{tabular}{ccccc}
  \hline
  Item & $\hat{\lambda}_{i0}$ & s.e. & $\hat{\lambda}_{i1}$ & s.e. \\
  \hline
     $y_1$ & -1.42 & (0.067) & 5.23 & (0.245) \\
    $y_2$ & \ 0.59 & (0.040) & 4.45 & (0.153) \\
    $y_3$ & \ 1.27 & (0.104) & 5.04 & (0.301) \\
    $y_4$ & \ 0.80 & (0.056) & 3.34 & (0.149) \\
  \hline
\end{tabular}
\end{table}

Table \ref{ab1} reports the threshold and loading estimates.
Corresponding standard errors have been obtained by $1000$ bootstrap
samples. We can observe that all the loadings are quite similar and
significant, confirming that there exists a latent variable common
to all of them, which summarizes the opinion pro/anti-abortion of
the respondents. This common factor is also used to classify the
different response patterns, since it is modelled as a mixture of
$k=2$ components. Table \ref{ab2} shows the clustering results of
the fitted model. For comparative purposes, classification obtained
by latent class analysis given by Bartholomew \textit{et al.} (2002)
and hierarchical clustering (HC) according to different methods
(complete linkage, single linkage and Ward method) is reported. As
already mentioned before, latent class analysis distinguishes
between two different behaviors of the respondents, those who tend
to be in favour of abortion (not conservative group), since they answer
'yes' at least to two out of the four items and those who are not in
favour of abortion (conservative group) since they reply yes to one
or none of the four items. On the contrary, if we look at the
hierarchical clustering, all the reported methods, a part from the
complete linkage one, suggest a more restrictive criterion of
classification, that is the conservative group is constituted only
by people who respond `no' to all the items. This classification is
also suggested by the proposed factor mixture analysis for
binary data (FMAB). Thus, the results obtained with FMAB are in agreement
with almost all the hierarchical clustering procedures but it is
more restrictive than the classical latent class analysis.

\begin{table}[ht]
\caption{\label{ab2} Class membership for the attitude to abortion data, according to different methods. Columns 2 and 3 are taken from Bartholomew \textit{et al.} (2002).}
\centering
\begin{tabular}{cccccc}
  \hline
  Response & LC & Complete & Single & Ward & FMAB \\
  pattern & allocation & HC & HC & HC & allocation \\
  \hline
    0000 & 1  & 1 &  1 & 1& 1\\
    0001 & 1  & 1 &  2 & 2& 2\\
    0010 & 1  & 2 &  2 & 2& 2\\
    0100 & 1  & 1 &  2 & 2& 2\\
    1000 & 1  & 1 &  2 & 2& 2\\
    0011 & 2  & 2 &  2 & 2& 2\\
    0101 & 2  & 1 &  2 & 2& 2\\
    0110 & 2  & 2 &  2 & 2& 2\\
    1100 & 2  & 1 &  2 & 2& 2\\
    0111 & 2  & 2 &  2 & 2& 2\\
    1011 & 2  & 2 &  2 & 2& 2\\
    1101 & 2  & 1 &  2 & 2& 2\\
    1110 & 2  & 2 &  2 & 2& 2\\
    1111 & 2  & 2 &  2 & 2& 2\\
  \hline
\end{tabular}
\end{table}

\subsection{Example 2: American students exposure to school and neighborhood violence}
This example has been extracted from the National Longitudinal
Survey of Freshmen (NLSF)\footnote{This research is based on data
from the National Longitudinal Survey of Freshmen, a project
designed by Douglas S. Massey and Camille Z. Charles and funded by
the Mellon Foundation and the Atlantic Philanthropies, website
http://nlsf.princeton.edu/.}. The NLSF aims at evaluating the academic
and social progress of American college students at regular
intervals in order to capture emergent psychological processes (by measuring
the degree of social integration and intellectual engagement) and to
control for pre-existing background differences with respect to
social, economic, and demographic characteristics. Data are
collected over a period of four waves (1999-2003). \\
For this analysis we have considered a part of the questionnaire
administered in the year 1999 that
 investigates the freshmen exposure to school and neighborhood
 violence at age six to ten.  It is composed by 21 binary items, 12 of them measuring violence in the schools,
 the remaining measuring violence in the neighborhood. The items are
 reported in Table \ref{item}.

\begin{table}[ht]
 \footnotesize{
\caption{\label{item} NLSF data: item description.}
 \begin{tabular}{ll}
   \hline
   Item & Question\\
   \hline
   w1q13a & In your grade school, when you were between the age of six and
   ten, did you see students fighting?    \\
   w1q13b & Students smoking? \\
   w1q13c & Student cutting class?      \\
   w1q13d & Students Cutting school?   \\
   w1q13e & Students verbally abusing teacher's?                                                              \\
   w1q13f & Did you see physical violence directed at teachers by students?                                   \\
   w1q13g & Vandalism of school or personal property?                                                         \\
   w1q13h & Theft of school or personal property?                                                              \\
   w1q13i & Students consuming alcohol?                                                                  \\
   w1q13j & Students taking illegal drugs?                                                               \\
   w1q13k & Students carrying knives as weapons?                                                               \\
   w1q13l & Students with guns?                                                                                \\
   w1q14a & In your neighborhood, before you were ten, do you remember seeing homeless people on the street?  \\
   w1q14b & Prostitutes on the street?                                                                         \\
   w1q14c & Gang members hanging on the street?                                                                \\
   w1q14d & Drug paraphernalia on the street?                                                                 \\
   w1q14e & People selling illegal drugs in public?                                                          \\
   w1q14f & People using illegal drugs in public?                                                            \\
   w1q14g & People drinking or drunk in public?                                                      \\
   w1q14h & Physical violence in public?                                                               \\
   w1q14i & Hearing the sound of gunshots?      \\  \hline
 \end{tabular}}
 \end{table}

\normalsize

Possible responses are `yes' (coded by 1) or `no' (coded by 0). The original sample was
 3924 students of different races. Since we consider several specifications of the proposed latent variable model we reduced computational time by analyzing a random subsample of 400 individuals.\\
We started with estimating different FMAB models with $q=1$ and $k$
ranging from $1$ to $4$. In all cases the one-factor model is
rejected according to both $GF$  and $LR$ test. Also looking at the
bivariate residuals there are several pairs of items that present
high values, confirming a bad fit. Thus we considered $q=2$ and
$k=1,2,3,4$. For these models $GF$ and $LR$ test still indicate that
the two-factor model is a poor fit to the data (Table \ref{fit}),
but the bivariate residuals lead to different conclusions. In Table
\ref{marg} the greatest bivariate residual ($Gffit$) for each pair
of responses, for chosen groups $k=3$, are shown (as in the
simulation study,  the number of groups has been determined
according to the AIC criterion, Table \ref{fit}). We can observe
that only one pair of items presents a residual equal to 5.16,
indicating that the two latent variables accounts for the pairwise
associations and thus that the fit is satisfactory. Evidently, the
classical overall goodness of fit tests are strongly affected by
sparseness present in the data and this leads to a wrong rejection
of the two-factor model.

 In the first four columns of
Table \ref{4} we reported the loading estimates with
associated standard errors in brackets, computed by 1000 bootstrap
samples. All the loadings are significant and most of them are
negative. We can observe that the items concerning violence in the schools present
high negative loadings related to the first factor, whereas the items
measuring violence in the neighborhood have high negative loadings
in correspondence of the second factor. Thus we can interpret the
first factor as ``\emph{absence of violence in the schools}". The
items that strongly influence it are those expressing cutting school
(w1q13c and w1q13d) and taking illegal drugs (w1q13j).

On the other hand the second factor can be interpreted as ``\emph{absence
of violence in the neighborhood}". It is interesting to notice that
also in this case the items that present the highest loadings are
w1q14d and w1q14e, that are the items related to the use and to the
traffic of illegal drugs.

 \begin{table}[ht]
 \caption{\label{fit} NLSF data: BIC, AIC, GF and LR for the two-factor model.}
\centering
\begin{tabular}{lccccccc}
  \hline
  $k$ & logL & $\sharp par$ &  BIC & AIC& GF & LR& $df$\\
    \hline
    1 &   -2658.257  & 62  &5687.99 & 5440.51 &1007733  &  904.84 & 139  \\
    2 &   -2649.596  & 68  &5706.61 & 5435.19 &323533.9  & 771.76 & 93  \\
    3 &  -2642.207  & 74   &5727.78 & 5432.41 &786993.6  & 764.37&  81   \\
    4 & -2643.952  & 80    &5767.22 & 5447.90 & 472935.8  &  766.12& 69  \\
  \hline
\end{tabular}
\end{table}

\begin{table}[ht]
\caption{\label{marg} NLSF data: greatest bivariate residuals for each response, $q=2$, $k=3$.}
\centering
\begin{tabular}{lcc}
  \hline
Response &Items& $Gffit$\\
  \hline
(0,0) & 7,8 & 0.59 \\
(0,1) & 13,14 & 3.95\\
(1,0) & 4,12  & 2.48\\
(1,1) & 8,15 & 5.16\\
  \hline
\end{tabular}
\end{table}

\begin{table}[ht]
\caption{\label{4} NLSF data: factor loadings with standard errors in brackets: unrotated and rotated solutions.}
\centering
\begin{tabular}{lcccc}
\hline
 \ \ \ Items \ \ \ & \ \ \ $\hat{\lambda}_{i1}$ \ & \ s.e. \ \ \ \ & \ \ \ $\hat{\lambda}_{i2}$ \ & \ s.e. \ \ \ \ \\
  \hline
 \ w1q13a   & -1.62 &(0.01) & \ 0.00  &(0.00) \\
 \ w1q13b & -2.18 &(0.04) & \ 0.03  &(0.01) \\
 \ w1q13c & -4.74 &(0.22) & \ 0.18  &(0.02) \\
 \ w1q13d & -5.76 &(0.26) & \ 0.43  &(0.03 \\
 \ w1q13e & -1.74 &(0.01) &-0.06 &(0.01)\\
 \ w1q13f & -1.12 &(0.01) &-0.25 &(0.01) \\
 \ w1q13g & -1.59 &(0.01) &-0.37 &(0.01) \\
 \ w1q13h & -1.45 &(0.01) &-0.10 &(0.01) \\
 \ w1q13i & -1.52 &(0.88) &-1.56 &(0.54) \\
 \ w1q13j & -8.66 &(1.73) & \ 1.72  &(0.52) \\
 \ w1q13k & -1.91 &(0.02) &-0.23 &(0.01) \\
 \ w1q13l & -0.81 &(0.08) &-0.85 &(0.08) \\
 \ w1q14a & -0.75 &(0.01) &-2.06 &(0.01) \\
 \ w1q14b & -0.78 &(0.01) &-2.00 &(0.02) \\
 \ w1q14c & -1.36 &(0.02) &-3.10 &(0.06) \\
 \ w1q14d & -2.41 &(0.05) &-5.08 &(0.11) \\
 \ w1q14e & -3.58 &(0.27) &-6.37 &(0.45) \\
 \ w1q14f & -2.44 &(0.04) &-3.39 &(0.09) \\
 \ w1q14g & -1.93 &(0.02) &-2.62 &(0.02) \\
 \ w1q14h & -2.08 &(0.01) &-2.34 &(0.01) \\
 \ w1q14i & -0.85 &(0.01) &-1.56 &(0.01) \\
   \hline
\end{tabular}
\end{table}

\begin{figure}
  \includegraphics[width=0.8\hsize,height=0.8\hsize]{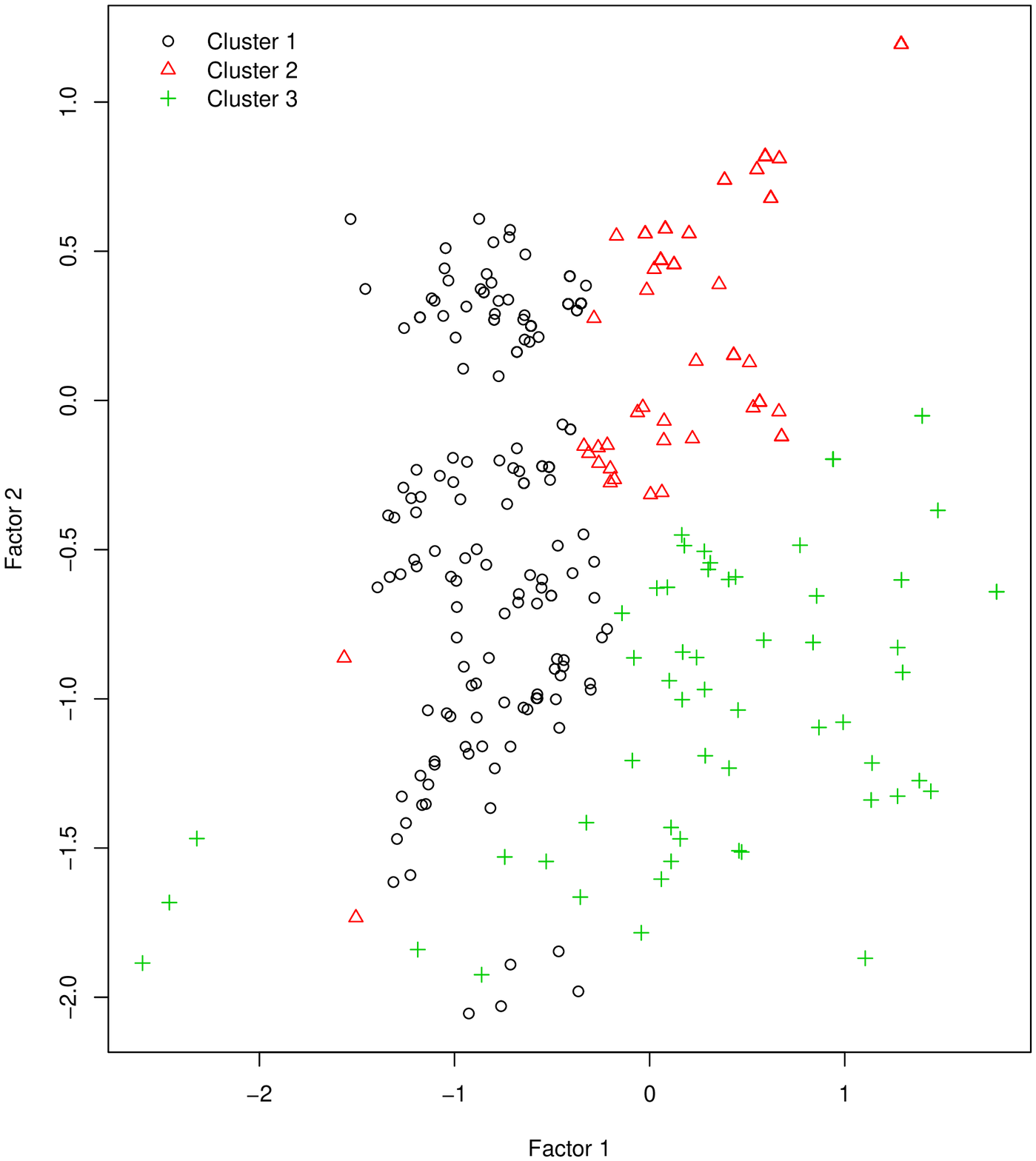}\\
  \caption{NLSF data: scatterplot of the estimated factor scores distinguished by group.}
  \label{f1}
\end{figure}

Figure \ref{f1} shows the scatterplot of the estimated factor scores
distinguished by group. The first cluster, drawn by circles, is
constituted by students who present low factor scores for the first
latent trait. In other words they are individuals that attended
unsafe schools despite the level of safety in the neighborhood. The
second cluster, indicated with triangles, is composed by individuals
who lived and brought up in no violent environments (both schools
and neighborhoods). The third cluster, indicated with crosses, is
composed by students who attended schools with little violence but
lived in violent neighborhoods.

The three groups of students can be also interpreted by computing
the weighted loadings, $\boldsymbol \Lambda \boldsymbol \mu_i$,
within each cluster $i$ with $i=1,\ldots,k$, reported in Table \ref{post}. In the first
group, the weighted loadings are positive for all the items, but
higher for those related to school,  and therefore for these
individuals, on average, the probability of giving answer 1
(presence of violence) is greater than giving answer 0 (absence of
violence). The second group of students is characterized by negative
weighted loadings and therefore they more likely answer 0. The third
group of students is characterized by a contrast between the climate
and safety in school and neighborhoods, as previously observed.
\begin{table}[ht]
\caption{\label{post} NLSF data: weighted loadings within each cluster.}
\centering
\begin{tabular}{rrrr}
  \hline
 & Cluster 1 & Cluster 2 & Cluster 3 \\
  \hline
w1q13a & 1.29 & -0.81 & -0.40 \\
  w1q13b & 1.74 & -1.07 & -0.58 \\
  w1q13c & 3.80 & -2.30 & -1.38 \\
  w1q13d & 4.66 & -2.71 & -1.94 \\
  w1q13e & 1.37 & -0.89 & -0.35 \\
  w1q13f & 0.83 & -0.65 & 0.04 \\
  w1q13g & 1.17 & -0.92 & 0.07 \\
  w1q13h & 1.12 & -0.76 & -0.23 \\
  w1q13i & 0.85 & -1.32 & 1.58 \\
  w1q13j & 7.27 & -3.69 & -4.26 \\
  w1q13k & 1.46 & -1.03 & -0.18 \\
  w1q13l & 0.45 & -0.71 & 0.86 \\
  w1q14a & 0.12 & -1.12 & 2.38 \\
  w1q14b & 0.16 & -1.11 & 2.30 \\
  w1q14c & 0.37 & -1.80 & 3.53 \\
  w1q14d & 0.75 & -3.04 & 5.74 \\
  w1q14e & 1.38 & -4.09 & 7.06 \\
  w1q14f & 1.15 & -2.44 & 3.63 \\
  w1q14g & 0.93 & -1.91 & 2.79 \\
  w1q14h & 1.11 & -1.88 & 2.40 \\
  w1q14i & 0.32 & -0.99 & 1.74 \\
   \hline
\end{tabular}
\end{table}


\section{Discussion}
The proposed model combines two methodologies coming from different
traditions. Latent trait analysis arose with the aim of evaluating
general abilities in education field. Nowadays it represents one of
the widest methods to deal with dimension reduction for binary data.
On the other hand, Gaussian mixture models have been shown to
be a powerful tool for clustering in many applications.\\
The combination of these two approaches allows to both measure
latent factors and to detect potential groups of observations,
simultaneously. In particular, clustering is obtained in the
dimensionally reduced space defined by the latent traits.

For these reasons it can be viewed as a generalization of the
proposal by Uebersax and Grove (1993) by allowing heteroscedastic
and multivariate mixture components. A similar approach has been
also discussed by Muth\`{e}n and Asparouhov (2006) in the context of
Item Response Theory (IRT) mixture models. However, differently from our
proposal, they do not explicitly assume a semi-parametric
distributional form for the latent variables through a mixture of
multivariate Gaussians. In addiction we tried fitting a IRT mixture
model with \textsf{Mplus 5} (Muth\`{e}n and Muth\`{e}n, 2007) on
NLSF data with two factors and three classes with the aim of making
a comparison with our solution but the algorithm did not achieve
convergence.

Results obtained on real data seem to be promising. However, some aspects still need to be investigated. From a computational point of view, the use of a full information
maximum likelihood method becomes computationally intensive as the number of latent variables increases. A further challenging aspect for future research is related to the goodness of fit of the model. As
already highlighted, the classical tests can be rarely used due to
the presence of sparse data. For this reason we referred to the
bivariate margins that, although very used in the latent trait
analysis, are measures of fit rather than tests. In this sense
limited information tests proposed in literature (Reiser, 1996,
Maydeu-Olivares and Joe, 2005) could be extended to the proposed
model. The analysis could be furthermore extended by
considering mixed type of observed variables and, more generally, by
putting the model in the framework of generalized linear models.



\end{document}